\documentclass[12pt,journal,a4paper,onecolumn]{IEEEtran}
\usepackage{stmaryrd}
\usepackage{amsfonts}
\usepackage{bbm}
\usepackage{mathrsfs}
\usepackage{graphicx}
\usepackage{epsfig}
\usepackage{color}
\usepackage{url}

\ifCLASSINFOpdf

\else

\fi
\usepackage{amsmath}
\usepackage{amssymb}
\usepackage{amsbsy}
\usepackage{amstext}
\usepackage{amsopn}
\usepackage{amsthm}
\usepackage{amssymb}
\usepackage{eufrak}
\usepackage{cite}

\usepackage{fancybox}
\usepackage{dashbox}

\begin{document}

\title{Performance Analysis of Two-Step Bi-Directional Relaying with Multiple Antennas}

\author{Mahshad~Eslamifar, Woon Hau~Chin, Chau~Yuen and~Yong Liang~Guan
\thanks{M. Eslamifar and Y.L. Guan are with the Nanyang Technological University. W.H. Chin was with the Institute for Infocomm Research, Singapore, he is now with Toshiba Research Europe Limited, Bristol, United Kingdom. C. Yuen is with Singapore University of Technology and Design, his research is supported by the Singapore University Technology and Design (grant no. SUTD-ZJU/RES/02/2011).}}
\maketitle

\begin{abstract}

In this paper we study decode-and-forward multi-antenna relay
systems that achieve bi-directional communication in two time
slots. We investigate different downlink broadcast schemes which
employ binary or analog network coding at the relay. We also
analyze and compare their performances in terms of diversity order
and symbol error probability. It is shown that if exact downlink
channel state information is available at the relay, using analog
network coding in the form of multi-antenna maximal-ratio transmit
beamforming to precode the information vectors at the relay gives
the best performance. Then, we propose a Max-Min antenna selection
with binary network coding scheme that can approach this
performance with only partial channel state information.
\end{abstract}

\IEEEpeerreviewmaketitle

\section{Introduction}
\IEEEPARstart {B}{i-directional} communications via a relay is
often encountered in intra-cell, intra-hotspot, or more recently
intra-picocell communication.

\begin{figure}[!t]
    \centering
    \epsfig{file=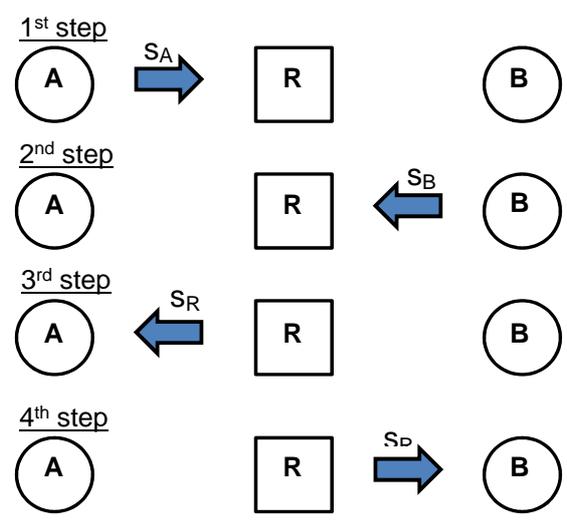, width=0.4\textwidth, clip=}
    \caption{Conventional four-step bi-directional relaying communication.}
    \label{fig:4step}
\end{figure}

\begin{figure}[!t]
   \centering
   \epsfig{file=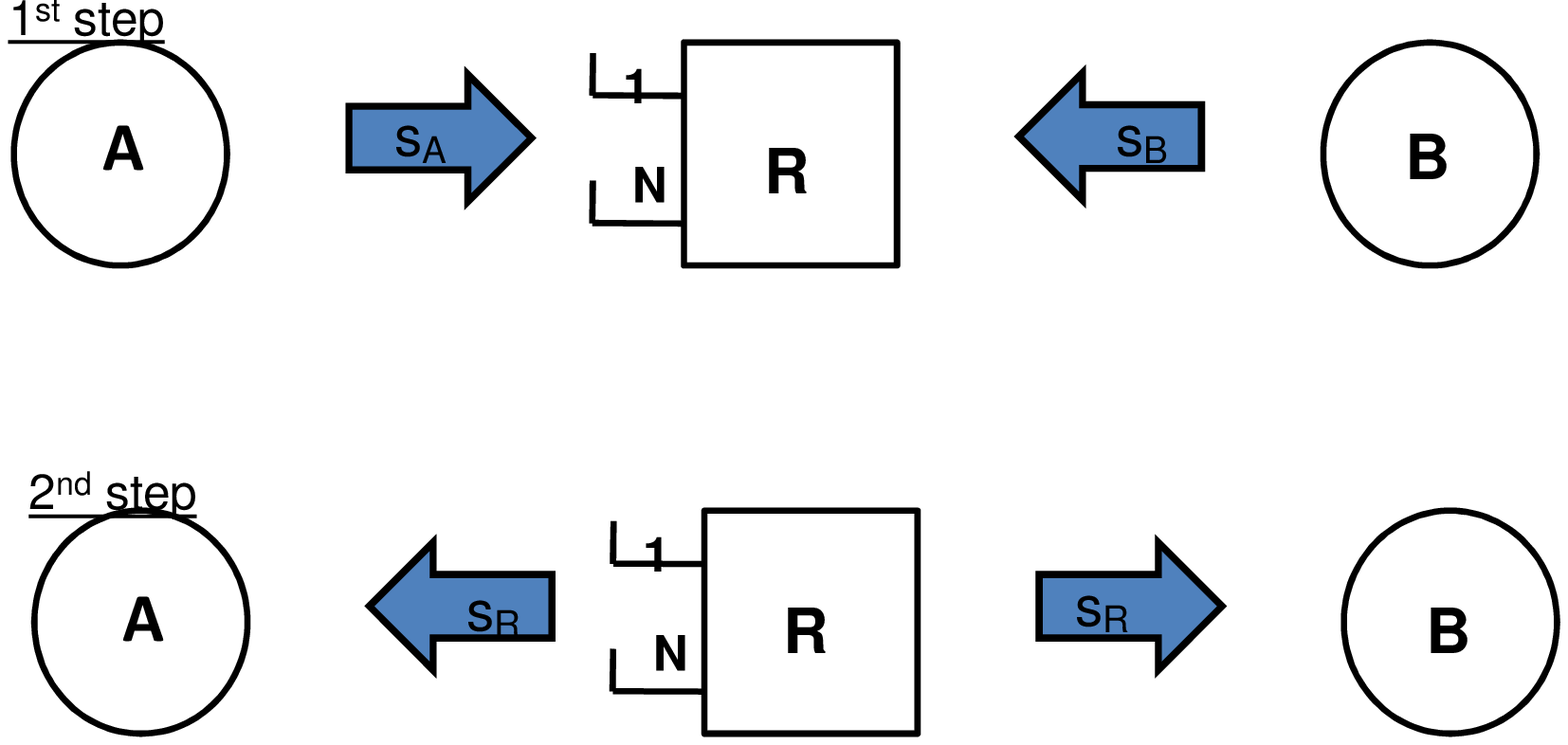, width=0.4\textwidth, clip=}
   \caption{Two-Step bi-directional communication with multi-antenna relay. Relay R broadcasts precoded versions of $s_R=\hat s_A \oplus \hat s_B$ or linear combination of the symbols $\hat s_A$ and $\hat s_B$ to nodes A and
B in the second time slot.}
   \label{fig:2step}
\end{figure}

Traditionally, bi-directional relaying requires 4 channel uses (or
4 steps) as communication in each direction requires 2 channel
uses, as shown in Fig.~\ref{fig:4step}. However, the efficiency of
the scheme can be improved by introducing network
coding~\cite{FNC} at the relay node. The relay node may perform
either analog network coding (ANC)~\cite{PNC,ANC,xor} or binary
network coding (BNC)~\cite{YCbasic,Popovski07}, while the
destination nodes perform self interference cancellation, to
reduce the channel uses of bi-directional relaying to 3 channel
uses (or 3 steps). In this paper, we consider a 2-step approach by
using two or more antennas at the relay, so that the relay node is
able to jointly detect both messages sent from the two source
nodes. The two messages are then combined using either binary
eXclusive-OR (XOR)~\cite{PNC,coded-tarokh} or analog beamforming.
This two-step procedure is illustrated in Fig.~\ref{fig:2step}.

In this paper, we study two different relaying schemes, ANC-based
Transmit Beamforming (TB) and BNC-based Space Time Block Coding
(STBC-BNC), and propose a novel BNC-based Max-Min Antenna Selection
(Max-Min AS-BNC) scheme. In the Max-Min AS-BNC scheme, we select an
antenna that has the largest channel gain among the worst channels
from each antenna to either nodes A or B. We analyze the diversity
gain and symbol error probability for the various multi-antenna
relaying schemes to get an insight into their relative system
performance and to compare their implementation requirements such as
the need for downlink ($2^{nd}$ step in Fig.~\ref{fig:2step})
channel state information (CSI), \emph{i.e} phase and amplitude of
the channels between each relay antennas and each node (A and B). It
is shown that Max-Min AS-BNC achieves full diversity order with the
need of coarse downlink CSI, \emph{i.e.} the relative amplitude of
the downlink channel, and approaches the performance of Transmit
Beamforming (TB) based on maximal ratio AF transmission which
requires exact CSI, and performs better than Space Time Block Coding
with Binary Network Coding (STBC-BNC) which requires no CSI.

In this paper, bold lower case and upper case letters denote vectors
and matrices, respectively; $[.]^T$ the transpose of a vector or a
matrix; $|.|$, the absolute value of a real number; $||.||$, the
norm of a vector and $Q(x)$, the $Q$-function of $x$ which is the
probability that a standard normal random variable will obtain a
value larger than $x$.

The paper is organized as follows. In the next section, we present the
system model for the different two-step two-way relaying schemes. In
Section III, we analyze the downlink channel for different
relaying schemes, in terms of symbol error probability and diversity
order. We present the numerical results and compare them with our
analytical results in Section IV. We then conclude our paper in
Section V.

\section{System Model}
We consider a network comprising of a pair of single-antenna nodes
(nodes A and B that would like to exchange information) and one
$N$-antenna relay node (node R). All nodes are half-duplex and
there is no direct link between nodes A and B. In the first
transmission interval, nodes A and B transmit $s_A$ and $s_B$
respectively to the relay R simultaneously and on the same
frequency. The multi-antenna relay R then uses maximum likelihood
(ML) detection to detect them jointly.

We assume that the probability distribution of all the uplink
channels and the downlink channels are the same, and the noise at
both nodes A and B is complex zero-mean additive white Gaussian with
variance $\sigma_S^2$. We will analyze the performance of the system
for the information flow from node B to node A, where the results
hold for the information flow from node A to node B by assuming that
the channels have the same distribution statistic.

In the first time slot (uplink), the relay R receives
\begin{equation}\label{eq:up1}
\textbf{y}_{R}=\textbf{H}_{up}\textbf{s}+\textbf{n}_{R}
\end{equation}
where $\textbf{s}=\left( s_{A}~s_{B} \right)^{T}$,
$\textbf{h}_{AR}=\left(h_{A1}\ldots h_{AN}\right)^{T}$,
$\textbf{h}_{BR}=\left( h_{B1}\ldots h_{BN}\right)^{T}$,
$\textbf{H}_{up}=\left(\textbf{h}_{AR}~\textbf{h}_{BR}\right)$, and
$\textbf{n}_{R}$ is a circularly symmetric complex Gaussian noise
vector
$\sim\mathcal{C}\mathcal{N}(\textbf{0}_N,\sigma^2_{R}\textbf{I}_N)$.
$h_{kj}$ is a Rayleigh flat fading channel with power mean
$E[|h_{kj}|^2]=\sigma_0^2$ between node $k$ and the $j^{th}$ relay
antenna where $k=$A or B and $j\in\{1,..,N\}$.

We assume that $\textbf{H}_{up}$ is known at the relay. So, the relay
performs maximum-likelihood decoding on the received vector to
obtain
\begin{equation}
{\bf{\hat s = }}\left( \hat s_A~\hat s_B \right)^{\rm{T}} = \arg
\min_{\bf{s}} || {{\bf{y}}_R  - {\bf{H}}_{up} {\bf{s}}} ||^2
\end{equation}
where $\hat s_A$ and $\hat s_B$ are the estimation of the
transmitted symbols by nodes $A$ and $B$.

In the second time slot (downlink), the relay uses different
multi-antenna precoding schemes to broadcast a message, that is formed by $\hat{s}_A$ and $\hat{s}_B$, to both nodes A and B. If the downlink
scheme is based on binary network coding (BNC), the relay performs
bit-wise XOR operation on the estimated symbols $\hat s_A$ and $\hat
s_B$ to obtain the network coded message
\[
s_{XOR} {\bf{ = }}\hat s_A  \oplus \hat s_B.
\]
Then, the relay broadcasts $\textbf{s}_R=\textrm{\it{f}}(s_{XOR})$,
which is a function of $s_{XOR}$. If the downlink scheme is based on
analoge network coding (ANC), $\textbf{s}_R$ is a linear combination
of the symbols $\hat s_A$ and $\hat s_B$,
$\textbf{s}_R=\textrm{\it{f}}(\hat{s}_A,~\hat{s}_B)$. In both cases,
$\textbf{s}_R$ is broadcasted by the relay to both nodes A and B in
the second time slot, as shown in Fig.~\ref{fig:2step}.

We denote the downlink channel vectors from the relay node to nodes
A and B by $\textbf{h}_{RA}=\left(h_{1A}\ldots h_{NA}\right)^T$ and
$\textbf{h}_{RB}=\left(h_{1B}\ldots h_{NB}\right)^T$ respectively.
Without loss of generality, $h_{jk}$ is a Rayleigh flat fading
channel with the same mean power as the uplink channels ,
\emph{i.e.} $E[|h_{jk}|^2]=\sigma_0^2$ between the $j^{th}$ relay
antenna and node $k$ where $k=$A or B and $j\in\{1,..,N\}$. The
average SNR at the relay node and at node A or B are
$\zeta_R=\frac{\mathcal {P}}{\sigma_R^2}$ and
$\zeta_S=\frac{\mathcal {P}}{\sigma_S^2}$ respectively, where
$\mathcal {P}=E[| s_A |^2]=E[| s_B |^2]=E[|| \textbf{s}_R ||^2]$.

Node A receives
\begin{equation}\label{generalreceive}y_A=\textbf{h}_{RA}^T\textbf{s}_R+n_A\end{equation}

If the relaying scheme is based on BNC, after estimating $s_{XOR}$,
node A performs bit-wise XOR operation on the estimated $s_{XOR}$
and $s_{A}$ to estimate $s_{B}$. If the relaying scheme is based on ANC, after canceling off the
contribution of $s_{A}$ from the received signal, node A estimates
$s_{B}$.

We consider three different relaying schemes which require different amount of
downlink channel state information (CSI) at the relay:
\begin{itemize}
\item No CSI: STBC-BNC relaying scheme, \item Coarse CSI: Antenna selection relaying scheme (\emph{i.e.}Max-Min AS-BNC),
\item Exact CSI: Transmit beamforming (TB) relaying scheme (based on ANC).\end{itemize}
In the exact CSI case, the relay knows the amplitude and phase of
the downlink channels. In the coarse CSI case, the relay just needs
to know the relative amplitude of the downlink channels.  Although
CSI can be inferred from the uplink in TDD system, this is usually
not feasible due to non-symmetrical impairments in the transmit and
receive circuitry. Therefore, the proposed Max-Min antenna selection
that requires only coarse CSI is more practical than the transmit
beamforming that requires perfect CSI.

We will show later that Max-Min antenna selection with BNC can
achieve full diversity, and it approaches the performance of TB
relaying scheme asymptotically. In addition to the fact that Max-Min
AS-BNC just needs coarse CSI, another advantage of using Max-Min
AS-BNC relaying scheme over TB is that in a distributed network of
multiple-antenna relays, implementing Max-Min AS-BNC is much easier
than implementing TB. To implement Max-Min AS-BNC, a method based on
time is selected: each relay will start its own timer with an
initial value, inversely proportional to $g_k$. The relay with the
least initial value or the most $g_k$ gets zero earlier than the
other relays~\cite{Distribution}. In this case to implement TB
scheme we need much more complicated algorithm to select one or two
relay(s) to transmit ${\textbf{s}_R}$ which results in the most
instantaneous SNR at nodes A and B.

\section{Downlink Analysis}
In the following, we derive the diversity order and symbol error
probability of the different downlink channel schemes and hence of
the whole system. We denote an exponential random variable $D$ with
parameter $\lambda$ as one with $E[D]=1/\lambda$; a
Gamma-distributed random variable $T$ with parameters
$(\Omega,\theta)$ as one with $E[T]=\Omega\theta$ and
$var(T)=\Omega\theta^2$. Also, a random variable $Z$ which is
complex Gaussian-distributed with parameters $(\mu,\sigma^2)$ has
$E[Z]=\mu$ and $var(Z)=\sigma^2$.

We assume both nodes A and B have sufficient downlink
channel knowledge to perform decoding. We employ M-PSK constellation
and coherent detection at the receiver, so according
to~\cite{Simon&Aloini1998,Trung2008,Pawula,Craig}, we can write the
symbol error probability (SEP) for the downlink channels as
\begin{eqnarray}\label{eq:Pr}
P_{d}=E_{\gamma}[P_{dA}]=1/\pi\int_{0}^{\pi-\pi/M}\psi_{\gamma}(g_{MPSK}/\sin^2\theta)\,d\theta,\end{eqnarray}
where $P_{dA}$ is the SEP of the downlink channels at node A,
$g_{MPSK}=\sin^2(\pi/M)$, $\gamma$ is the instantaneous receive SNR
and $\psi_{\gamma}(t)$ is the moment generating function (MGF) of
$\gamma$. Note that the proposed schemes can be applied to any
modulation. We use the M-PSK to demonstrate how the analysis can be
done, and the extension to other constellations is straightforward.

According to~\cite{Trung2008}, the diversity order of the downlink
channel can be found by
\begin{equation}\label{eq:Gd}\lim_{\zeta_S\to
\infty}-\frac{\log(\psi_{\gamma}(g_{MPSK}))}{\log(\zeta_S)}.\end{equation}

\subsection{Max-Min Antenna Selection with Binary Network Coding (Max-Min AS-BNC)}\label{sec:maxmin}
In the Max-Min AS-BNC scheme, we select just one antenna of the
relay node to send $s_{XOR}$, at all times. The process of selection
is done in two steps. First, we select the worst downlink channel
$h_{ik_{i}}$ for every antenna $i\in\{1,...,N\}$, \emph{i.e.}
\begin{equation}\label{eq:step1}k_i=\arg\underset{r}\min | h_{ir}|~,~~~~r\in\{A,B\}\end{equation}

Then, the relay antenna $j$ with the best downlink channel among those in (\ref{eq:step1}) is selected, {\emph{i.e.}}
\begin{equation}\label{eq:step2}j=\arg\underset{i}\max |h_{ik_i}|~,~~~~i \in\{1,...,N\}.\end{equation}

This antenna $j$ at the relay node is then used to broadcast
$s_{XOR}$ to nodes A and B. So, node A receives
\begin{equation}\label{eq:maxmin1}y_A=h_{jA}s_{XOR}+n_A.\end{equation}

Hence, the instantaneous SNR, $\gamma$, at node A is
\begin{equation}\label{eq:maxmin2}\gamma=|h_{jA}|^2\zeta_S.
\end{equation}

In this relaying scheme, there are two states. State 1 is when
the channel magnitude from the selected antenna at the relay node
to node A is less than to node B, {\emph{i.e.}} $|h_{jA}| < |h_{jB}|$, or
\begin{equation}\min\{|h_{jA}|,|h_{jB}|\}=|h_{jA}|,\end{equation}
and State 2 is when the channel magnitude from the selected antenna
at the relay node to node A is more than to node B, \emph{i.e.}
$|h_{jA}|> |h_{jB}|$, or
\begin{equation}\min\{|h_{jA}|,|h_{jB}|\}=|h_{jB}|.\end{equation}

Hence, the SEP at node A is
\begin{equation}
\label{eq:sepmaxmin}\rm{SEP=SEP_1P_1+SEP_2P_2},
\end{equation}
where ${\rm SEP}_i$ is the SEP conditioned on State $i$ and ${\rm
P}_i$ is the probability of occurrence of State $i$. Since the
channel gain in State 1 is less than the channel gain in State 2,
the symbol error probability of State 1 will be higher than the
symbol error probability of State 2, {\emph{i.e.}}
$\rm{SEP_1}>\rm{SEP_2}$. Since all channels are i.i.d,
$\rm{P_1=P_2=1/2}$, the overall SEP will be upper-bounded by
$\rm{SEP_1}$ and lower-bounded by $\rm{SEP_1}/2$, {\emph{i.e.}}
\begin{equation}
\label{eq:maxmininequality}\rm{\frac{SEP_1}{2} \leq SEP \leq SEP_1}.
\end{equation}

To use the MGF to work out the diversity order and SEP of the
downlink channel, we need to know the PDF of $|h_{jA}|^2$ in State
1.

From~\cite{orderstatistics}, the cumulative density function~(CDF)
of $|h_{ik_{i}}|^2$ can be shown to be
\begin{equation}\label{maxminpdf1}F(|h_{ik_{i}}|^2)=1-\exp(-2|h_{ik_{i}}|^2/\sigma^2_0).\end{equation}
Hence, the PDF of $|h_{jA}|^2$ in State 1 is
\begin{eqnarray}\label{maxminpdf2}f(|h_{jA}|^2)=\frac{2N}{\sigma_0^2}\exp(-2|h_{jA}|^2/\sigma_0^2)(1-\exp(-2|h_{jA}|^2/\sigma_0^2))^{N-1}.\end{eqnarray}
Using binomial expansion, we have
\begin{eqnarray}\label{maxminMGF1}(1-\exp(-2|h_{jA}|^2/\sigma_0^2))^{N-1}
=\displaystyle\sum_{k=0}^{N-1}{N-1\choose
k}(-1)^k\exp(-2k|h_{jA}|^2/\sigma_0^2).\end{eqnarray}

The MGF of the instantaneous SNR is thus
\begin{eqnarray}\label{maxminMGF2}\psi_\gamma(t)&=&E[\exp(-t\gamma)]\nonumber\\
&=&\int_0^{\infty}\exp(-t|h_{jA}|^2\zeta_S)f(|h_{jA}|^2)\,d
|h_{jA}|^2.\end{eqnarray} Substituting (\ref{maxminMGF1}) into
(\ref{maxminpdf2}) and then using (\ref{maxminMGF2}), the MGF of the
instantaneous SNR in the Max-Min AS-BNC scheme in State 1 is
\begin{eqnarray}\psi_\gamma(t)&=&\sum_{k=0}^{N-1}{N-1\choose
k}\frac{2N(-1)^k}{\sigma_0^2}\int_0^{\infty}\exp(-|h_{jA}|^2(t\zeta_S+2(k+1)/\sigma_0^2)\,d
|h_{jA}|^2\nonumber\\\label{eq:maxminMGF4}
&=&\sum_{k=0}^{N-1}{N-1\choose
k}\frac{N(-1)^k}{1+k+t\sigma^2_0\zeta_S/2}\\\label{eq:maxminMGF3}
&=&\frac{N!}{\prod_{k=0}^{N-1}(k+1+t\sigma_0^2\zeta_S/2)}.\end{eqnarray}
The equality between (\ref{eq:maxminMGF4}) and (\ref{eq:maxminMGF3})
can be easily shown using partial fraction decomposition~\cite{PFD}.

According to (\ref{eq:Gd}), the diversity order of the downlink
channel of the Max-Min AS-BNC scheme in State 1 is
\begin{eqnarray}\label{eq:maxmin4}G_d&=&\lim_{\zeta_S \to \infty} -\frac{\log N!}{\log
\zeta_S}+\lim_{\zeta_S \to
\infty}\frac{\displaystyle\sum_{k=0}^{N-1} \log
(k+1+g_{MPSK}\sigma_0^2\zeta_S/2)}{\log \zeta_S}\nonumber\\
&=&0+N=N.
\end{eqnarray}
Following the inequalities in (\ref{eq:maxmininequality}), we
conclude that the Max-Min AS-BNC relaying scheme is able to achieve
the full diversity order of $N$ for the downlink channel and also
for the whole system.

Similarly, using (\ref{eq:maxminMGF3}), the SEP of the downlink
channel can be lower-bounded as
\begin{equation}\label{eq:maxmin5}P_d^{Max-Min} = E_h[P_{eA}]\geq\frac{1}{2\pi}\int_0^{\pi-\pi/M}\psi_\gamma\left(\frac{g_{MPSK}}{\sin^2\theta}\right)\,d\theta.\end{equation}

\subsection{Transmit Beamforming (TB)}
In the TB downlink relaying scheme, the detected uplink data
$\hat{s}_A$ and $\hat{s}_B$  are precoded using analog transmission
weight vectors $\textbf{v}_B$ and $\textbf{v}_A$ respectively based
on the principles of maximum ratio transmission (MRT)~\cite{MRT}
\begin{equation}\label{eq:TB1}\textbf{v}_A=\frac{\textbf{h}^*_{RA}}{||\textbf{h}_{RA}||}, \textbf{v}_B=\frac{\textbf{h}^*_{RB}}{||\textbf{h}_{RB}||}.\end{equation}

The relay R then sends
\begin{equation}\label{eq:SVD1}\textbf{s}_R=\frac{\textbf{v}_B\hat{s}_A+\textbf{v}_A\hat{s}_B}{\sqrt2}.\end{equation}
where $\sqrt 2$ is to normalize the transmission power at the relay
node. The receiver output at node A, after subtracting
$\textbf{h}^T_{RA}\textbf{v}_Bs_A$ from the received vector is
\begin{eqnarray}\label{eq:SVD2}y_A=\frac{\textbf{h}^T_{RA}\textbf{v}_A\hat s_B}{\sqrt2}+n_A
=\frac{||\textbf{h}_{RA}||\hat s_B}{\sqrt2}+n_A.\end{eqnarray} The
instantaneous SNR ($\gamma$) at node A will be
\begin{equation}\label{eq:SVD3}
\gamma=\frac{ ||\textbf{h}_{RA}||^2\zeta_S}{2}.\end{equation}

Since we assume perfect cancellation at the receiver, the downlink
channel is equivalent to a point-to-point MISO channel. The optimal
precoding scheme for point-to-point transmission is MRT. This
explains our choice of (\ref{eq:TB1}) as the TB weight vectors.
Since TB linearly sums $\hat{s}_A$ and $\hat{s}_B$, it is a form of
ANC (analog network coding). From~\cite{larssonmimobook}, the
diversity order of MRT is $N$, therefore the diversity order of the
downlink channel and the entire bi-directional relaying system is
$N$.

Since $||\textbf{h}_{RA}||^2=\sum_{i=1}^N\mid h_{iA}\mid^2$ is the
sum of $N$ independent exponentially distributed random variables
each with $\lambda=1/\sigma^2_0$,  $||\textbf{h}_{RA}||^2$ is a
Gamma random variable with parameters $(N,\sigma_0^2)$. Hence, the
MGF of $\gamma$ is
\begin{equation}\label{eq:SVD4}\psi_\gamma(t)=\left(\frac{\lambda}{\lambda+t\zeta_S/2}\right)^N=\left(\frac{1}{1+t\sigma_0^2\zeta_S/2}\right)^N,\end{equation}

Using (\ref{eq:SVD4}) and (\ref{eq:Pr}), the SEP can be written as
\begin{eqnarray}\label{eq:SVD6}P_d^{TB}=E_h[P_{dA}]=\frac{1}{\pi}\int_0^{\pi-\pi/M}\psi_\gamma\left(\frac{g_{MPSK}}{\sin^2\theta}\right)\,d\theta.\end{eqnarray}

\subsection{Space-Time Block Coding with Binary Network Coding (STBC-BNC)}
For the STBC-BNC scheme, at the relay node, binary network coded
symbols at times $n,...,n+N$ are buffered and encoded using STBC to
produce $\textbf{s}_R$. Alamouti coding is an orthogonal STBC for
two transmit antennas, which achieves full diversity order without
sacrificing transmission rate. In the Alamouti-BNC scheme, the
downlink signal $\textbf{s}_R$ is
\[\textbf{s}_R=\left(
\begin{array}{cc}
s_{XOR}(n)&-s_{XOR}^*(n+1)\\
s_{XOR}(n+1)&s_{XOR}^*(n)\\
\end{array}
\right).\]

Correspondingly, node A receives
\begin{equation}\label{eq:alamouti2}\left(
\begin{array}{cc}
y_A(n)&y_A(n+1)\\
\end{array}
\right)=\frac{\textbf{h}^T_{RA}\textbf{s}_R}{\sqrt2}+n_A.\end{equation}
In general, by using orthogonal STBC at an $N$-antenna relay, it is
easy to show that the instantaneous SNR ($\gamma$) received at the
destination node is
\begin{equation}\label{eq:alamouti3}\gamma=\frac{|| \textbf{h}_{RA} ||^2\zeta_S}{N},\end{equation}
which is the same as (\ref{eq:SVD3}) for the TB relaying scheme
except for the denominator of $N$ instead of 2. Hence, we conclude
that the diversity order of the downlink channel and also the whole
system, is $N$ for the STBC-BNC relaying scheme.

For the same reason, the MGF of $\gamma$ is the same as
(~\ref{eq:SVD4}) with $\zeta_S$ changed to $\zeta_S/N$
\begin{equation}\label{eq:alamouti4}
\psi_\gamma(t)=\left(\frac{1/\sigma_0^2}{1/\sigma_0^2+t\zeta_S/N}\right)^N.
\end{equation}

Using (\ref{eq:alamouti4}) and (\ref{eq:Pr}), the SEP can be written
as
\begin{eqnarray}\label{eq:alamouti5}P_d^{\rm STBC-BNC}=E_h[P_{dA}]=\frac{1}{\pi}\int_0^{\pi-\pi/M}\psi_\gamma\left(\frac{g_{MPSK}}{\sin^2\theta}\right)\,d\theta.\end{eqnarray}

\section{Simulation and Discussion}
\subsection{Diversity and SEP Comparison}
We approximate the overall performance by assuming that overall
bi-directional communication will be erroneous if an error event
occurs in either the uplink or the downlink. Hence,
\begin{equation}\label{eq:tot1}P_{tot}\approx P_u+P_d(1-P_u),\end{equation}
where $P_u,P_d$ and $P_{tot}$ are SEP of the uplink channel,
downlink channel and the whole system respectively. Since $P_uP_d\ll
1$, (\ref{eq:tot1}) can be simplified to
\begin{equation}\label{eq:tot2}P_{tot}\approx P_u+P_d.\end{equation}
The union bound property~\cite{proakis} is used to compute an upper bound on $P_u$,
\begin{equation}\label{eq:up4}P_{u} \leq \sum_{\mathcal
{S}\neq \mathcal
{S}_0}Q\left(\sqrt{\frac{\parallel(\textbf{S}_0-\textbf{S})\textbf{h}\parallel^{2}}{2\sigma_{R}^{2}}}\right),
\end{equation}
while the SEP derived for the different downlink schemes in
(\ref{eq:maxmin5}), (\ref{eq:SVD6}) and (\ref{eq:alamouti5}) will be
used as $P_{d}$ in (\ref{eq:tot2}).


In Fig.~\ref{fig:up+downNC}, we plot the analytical and simulated
SEP of the ANC and BNC-based relaying schemes. Close agreement
between the analytical result and the simulation result,
especially in the high SNR region, is observed. Also, TB has an
advantage of 0.5 dB over the BNC-based schemes. Interestingly, the
Max-Min AS-BNC scheme and Alamouti-BNC scheme have very close
performance at all SNRs.

\begin{figure}[!t]
    \centering
    \epsfig{file=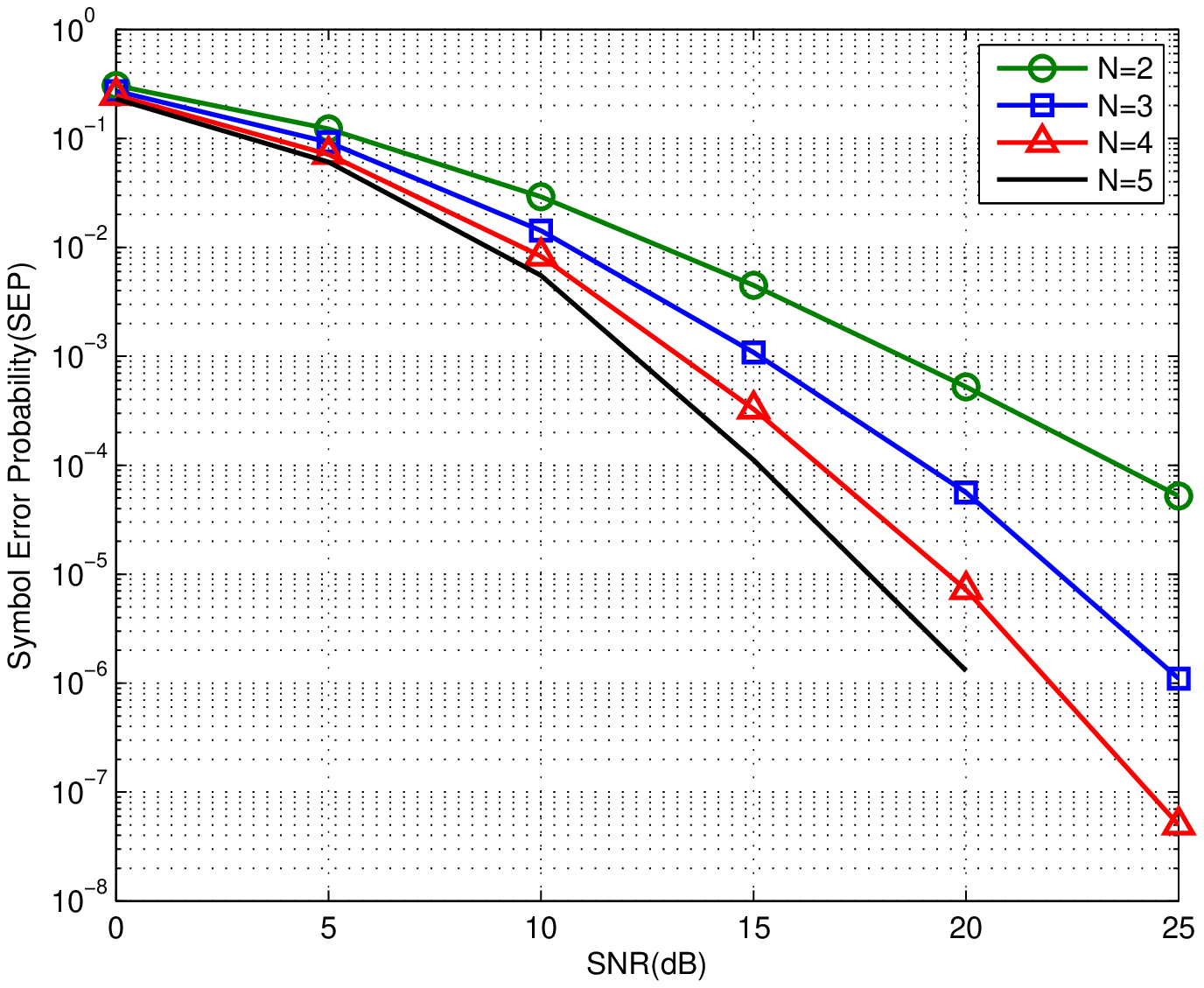,width=0.6\textwidth,clip=}
    \caption{Simulation results of downlink (2nd step) of Max-Min AS-BNC with different relay antenna number $N$.}
    \label{fig:maxmindl}
\end{figure}

\begin{figure}[!t]
    \centering
    \epsfig{file=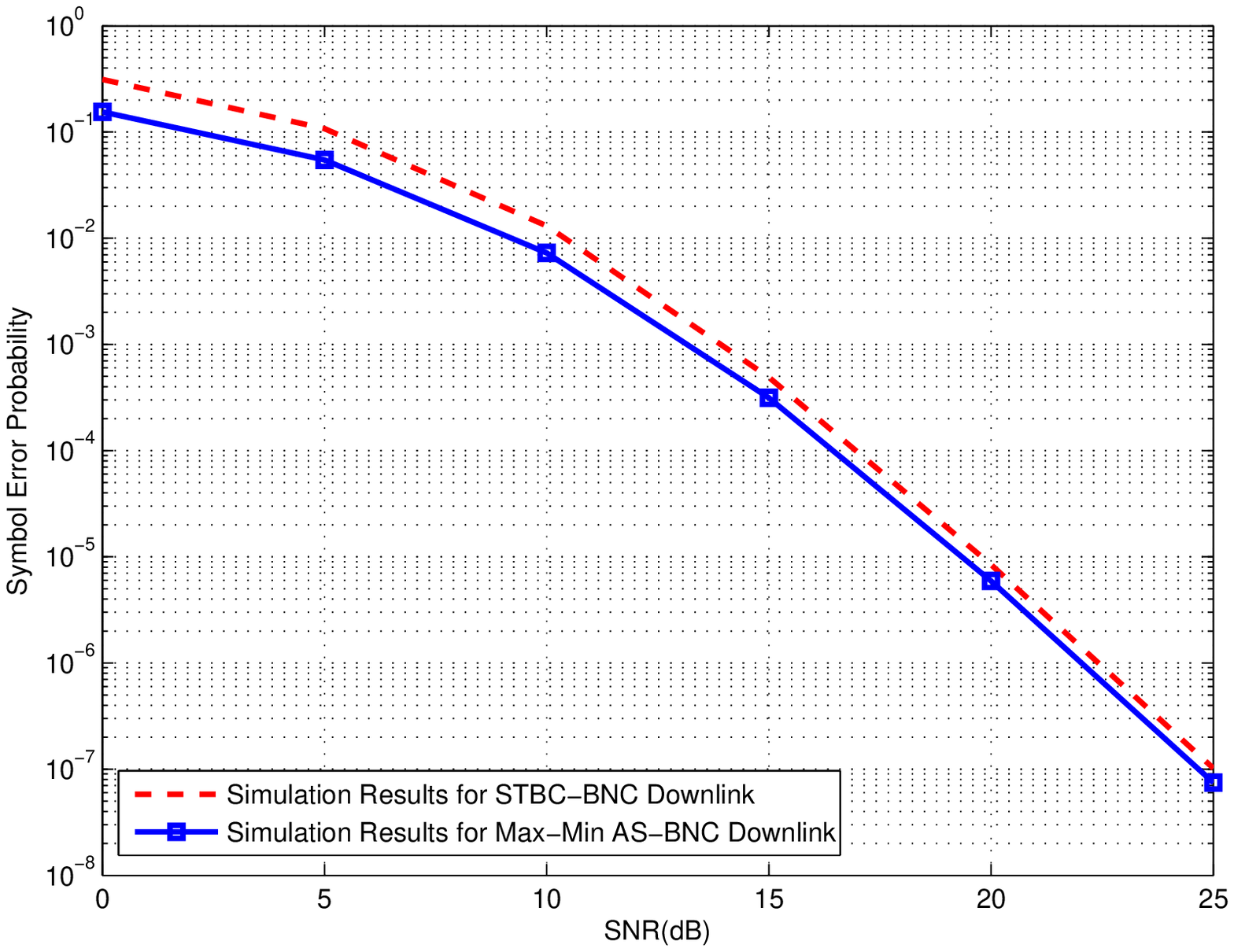,width=0.6\textwidth,clip=}
    \caption{Analytical SEP of two-way relaying using QPSK with Max-Min AS-BNC and STBC-BNC for four-antenna relay.}
    \label{fig:downNC4ant}
\end{figure}

In order to verify our analysis of the diversity order of Max-Min
AS-BNC, we show the simulation results of the downlink portion
with various antenna configuration in Fig.~\ref{fig:maxmindl}.
From the figure, we can corroborate that the scheme achieves full
downlink diversity order. Since the uplink performs ML decoding,
we may conclude that the scheme achieves full end-to-end diversity
order too. Additionally, we compare the Max-Min AS-BNC and
STBC-BNC schemes by simulation for four-antenna relay in
Fig.~\ref{fig:downNC4ant}. It shows that the Max-Min AS-BNC scheme
has a coding gain over the other. This advantage will be analyzed
in the next section.

\subsection{Asymptotic Analysis}
We analyze the asymptotic SEP ratio of downlink relaying scheme A
relative to scheme B as:
$\frac{SEP_A}{SEP_B}\mid_{SNR\rightarrow\infty}$. If this ratio is
less than one, the relaying scheme A is deemed to have SEP
performance gain over the relaying scheme B asymptotically.

We begin with the SEP ratio of TB over Max-Min AS-BNC
\begin{eqnarray}\label{proof:tablethree1}\frac{SEP_{TB}}{SEP_{Max-Min~AS-BNC}}|_{SNR \to \infty}\nonumber&=&\lim_{\zeta_S \to \infty}
\frac{\int_0^{\frac{\pi}{M}}\frac{2\sin^2\theta}{(\sin^2\theta+g_{MPSK}\sigma_0^2\zeta_S/2)^N}
d\theta}{\int_0^{\frac{\pi}{M}}\frac{N!\sin^2\theta}{\prod_{k=0}^{N-1}(\sin^2\theta(k+1)+g_{MPSK}\sigma_0^2\zeta_S/2)}\,d\theta}
\\\nonumber&=& \lim_{\zeta_S \to
\infty}\frac{\int_0^{\frac{\pi}{M}}\frac{2\sin^2\theta}{(g_{MPSK}\sigma_0^2\zeta_S/2)^N}
\,d\theta}{\int_0^{\frac{\pi}{M}}\frac{N!\sin^2\theta}{\prod_{k=0}^{N-1}(g_{MPSK}\sigma_0^2\zeta_S/2)}\,d\theta}
\\\nonumber&=& \lim_{\zeta_S \to
\infty}\frac{(g_{MPSK}\sigma_0^2\zeta_S/2)^N\int_0^{\frac{\pi}{M}}2\sin^2\theta
\,d\theta}{(g_{MPSK}\sigma_0^2\zeta_S/2)^NN!\int_0^{\frac{\pi}{M}}\sin^2\theta
\,d\theta}\\\label{eq:TBtoMaxMin}&=&\frac{2}{N!}.
\end{eqnarray}

Next, the SEP ratio of STBC-BNC over Max-Min AS-BNC is
\begin{eqnarray}\label{proof:tablethree2}\frac{SEP_{STBC-BNC}}{SEP_{Max-Min~AS-BNC}}|_{SNR \to \infty}\nonumber&=&\lim_{\zeta_S \to \infty}
\frac{\int_0^{\frac{\pi}{M}}\frac{2\sin^2\theta}{(\sin^2\theta+g_{MPSK}\sigma_0^2\zeta_S/N)^N}
d\theta}{\int_0^{\frac{\pi}{M}}\frac{N!\sin^2\theta}{\prod_{k=0}^{N-1}(\sin^2\theta(k+1)+g_{MPSK}\sigma_0^2\zeta_S/2)}\,d\theta}
\\\nonumber&=& \lim_{\zeta_S \to
\infty}\frac{\int_0^{\frac{\pi}{M}}\frac{2\sin^2\theta}{(g_{MPSK}\sigma_0^2\zeta_S/N)^N}
\,d\theta}{\int_0^{\frac{\pi}{M}}\frac{N!\sin^2\theta}{\prod_{k=0}^{N-1}(g_{MPSK}\sigma_0^2\zeta_S/2)}\,d\theta}
\\\nonumber&=& \lim_{\zeta_S \to
\infty}\frac{(g_{MPSK}\sigma_0^2\zeta_S/N)^N\int_0^{\frac{\pi}{M}}2\sin^2\theta
\,d\theta}{(g_{MPSK}\sigma_0^2\zeta_S/2)^NN!\int_0^{\frac{\pi}{M}}\sin^2\theta
\,d\theta}\\\label{eq:STBCtoMaxMin}&=&\frac{N^N}{2^{N-1}N!}.
\end{eqnarray}

\begin{figure}[!t]
    \centering
    \epsfig{file=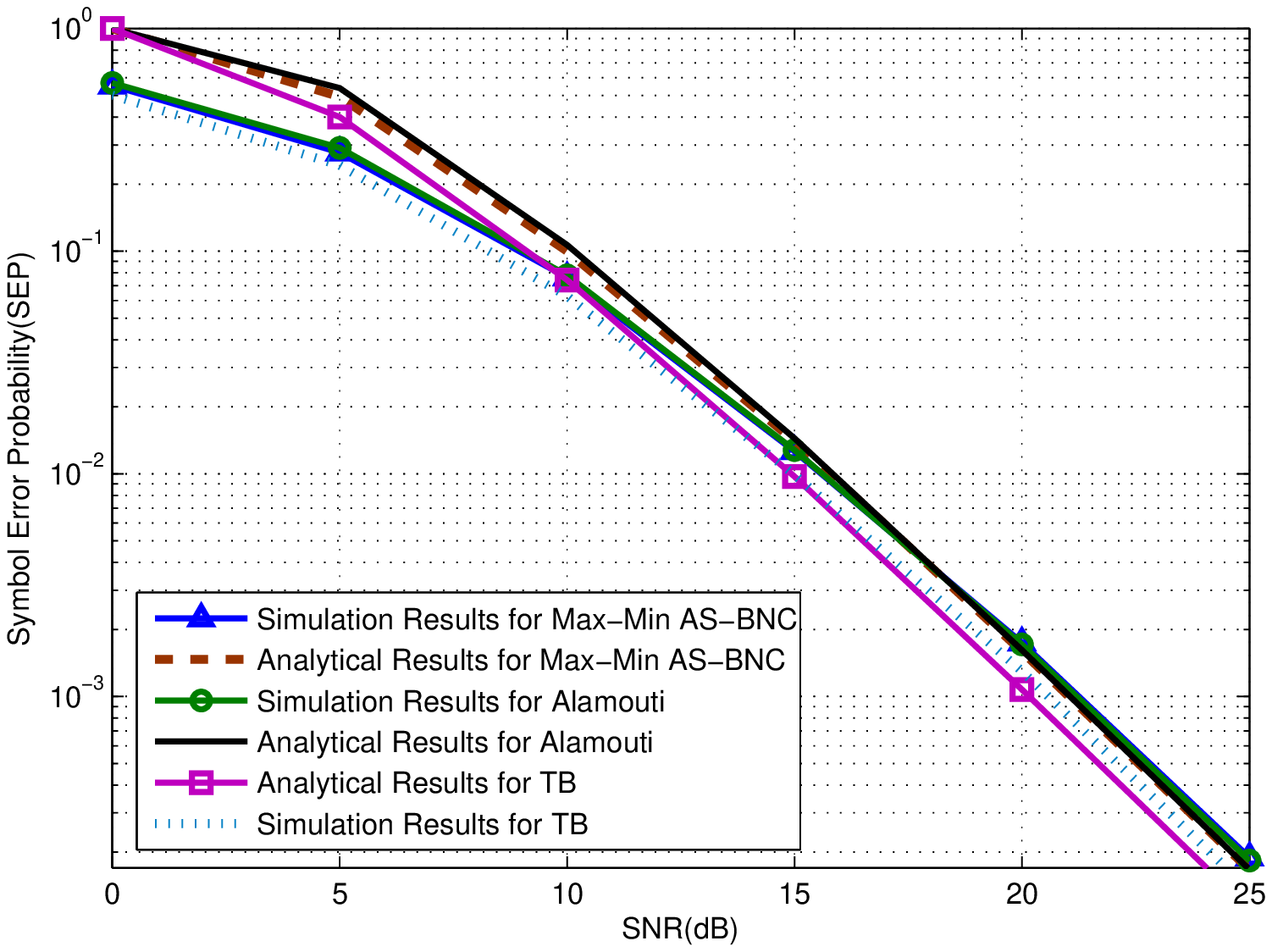,width=0.6\textwidth,clip=}
    \caption{Simulation and Analytical SEP of two-way relaying using QPSK with TB, Max-Min AS-BNC and Alamouti-BNC.}
    \label{fig:up+downNC}
\end{figure}

(\ref{eq:TBtoMaxMin}) and (\ref{eq:STBCtoMaxMin}) suggest that
with two antennas at the relay, all the relaying schemes have the
same downlink SEP performance, while with more than two antennas
at the relay, TB achieves the best downlink SEP performance,
followed by Max-Min AS-BNC, then STBC-BNC. These analysis explain
the observations in Fig.~\ref{fig:up+downNC} and
Fig.~\ref{fig:downNC4ant} respectively. Note that, although we use
the lower bound of the SEP of Max-Min AS-BNC in the SEP ratio
derivation, the simulation results in Fig.~\ref{fig:up+downNC}
have shown that the lower bound is very tight especially in the
high SNR region.

\section{Conclusion}

In this paper, we present and analyze three different multi-antenna
relaying schemes with binary or analog network coding at the relay
that achieve bi-directional wireless relay communication in two
steps. The relaying scheme with analog network coding employs
Transmit Beamforming (TB) with maximal ratio transmission. The
relaying schemes with binary network coding employ Max-Min Antenna
Selection (Max-Min AS-BNC) and Space-Time Block Coding (STBC-BNC).
We derive the diversity order and symbol error probability formulas
of these two-step bi-directional relaying schemes considering two
single-antenna nodes connected via a multi-antenna relay.

We show that while all the relaying schemes presented achieve full
diversity order, different schemes allow for different trade-offs
between complexity and performance. Specifically, TB performs best
in terms of SEP performance, but it needs exact downlink CSI at
the relay. Max-Min AS-BNC comes next and is able to tolerate
coarse CSI and can be implemented in a distributed multiple-relay
system, followed by STBC-BNC which does not require CSI.
Asymptotic SEP ratios are also derived to help evaluate the
relative performance of these full-diversity two-way relaying
schemes when the SNR is high.

\addcontentsline{toc}{chapter}{Bibliography}   
\bibliographystyle{ieeetr}
\bibliography{Refrence}

\clearpage

\end{document}